\documentclass[prb,10pt,aps,twocolumn,showpacs,showkeys]{revtex4-2}
\usepackage{graphicx}
\usepackage{soul}
\usepackage{color}
\usepackage{url}
\usepackage{lineno}
\usepackage{url}
\usepackage{hyperref}
\usepackage{xspace}
\usepackage{epsfig}
\usepackage{amssymb}
\usepackage{amsmath}
\usepackage{multirow}
\usepackage{amsfonts}
\usepackage{amsthm}
\usepackage{mathrsfs}
\usepackage{xcolor}
\usepackage{textcomp}
\begin{document}
\title{Asynchronous Breathers in Hamiltonian SQUID Metamaterials}
\author{N. Lazarides}
\affiliation{Academic Support Department, Abu Dhabi Polytechnic, Abu Dhabi,
             P.O. Box 111499, United Arab Emirates}
\date{\today}
\begin{abstract}
A one-dimensional SQUID (superconducting quantum interference device) array /
metamaterial is investigated numerically with respect to its localization 
properties due to nonlinearity in the absence of dissipation and periodic driving. 
The system possesses a conserved Hamiltonian function representing its energy, 
and supports localized modes of the discrete breather type even in the presence 
of a moderately high dc flux bias. The appearance of discrete breathers in that 
system has been largely overlooked in literature. We find a new type of discrete 
breather that is asynchronous, meaning that the frequency of oscillation of the 
SQUID at the central breather site is different than that of the SQUIDs at the 
other sites of the metamaterial. Nonlinear localization is investigated by
initializing the system with a single-site excitation of given amplitude (initial
amplitude) for a fixed value of the coupling coefficient, while parameters such 
as the dc flux bias, the single-site initial excitation amplitude, and/or the  
SQUID can vary independently. Using the energetic participation ratio as a 
measure of the degree of localization, the existence of asynchronous highly
localized modes and transitions between delocalized (extended) and localized 
modes are identified. 
\end{abstract}
\maketitle
\section{Introduction}  
\begin{figure}[!h]
\includegraphics[angle=0, width=0.85 \linewidth]{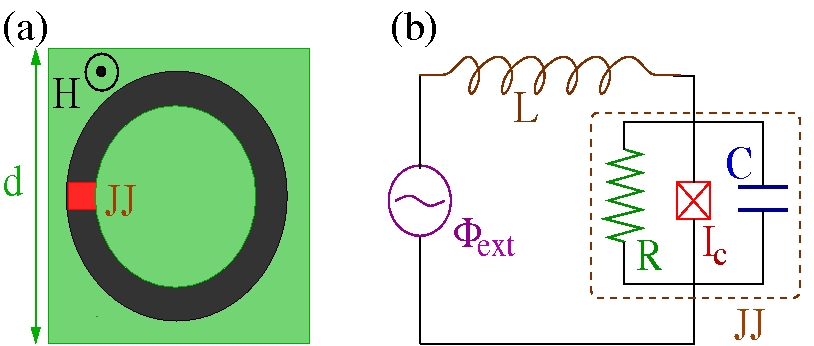} 
\caption{(Color online)
(a) Schematic of an rf SQUID in a field ${\bf H}$, which can be composed by dc 
    (constant) and/or ac (time-periodic) components, and 
(b) equivalent electrical circuit. 
    The real Josephson junction is represented by the circuit elements within the 
    brown-dashed box. 
}
\label{fig01}
\end{figure}

Periodic arrays of SQUIDs (superconducting quantum interference devices) in one 
and two dimensions form magnetic metamaterials \cite{Lazarides2007} that exhibit 
several extraordinary properties described in recently published reviews
\cite{Jung2014,Anlage2011,Lazarides2018b}.  The SQUID, the elementary unit of 
these metamaterials, is a highly nonlinear oscillator with a strong resonant 
response to applied magnetic fields; its simplest version can be realized by a 
superconducting ring which is interrupted by a Josephson junction 
\cite{Josephson1962}, which can be regarded as a nonlinear inductance in a 
superconducting circuit; a schematic of such a device in a perpendicular,
time-periodic magnetic field and its equivalent electrical circuit is shown in 
Fig. \ref{fig01}. SQUID systems often exhibit very rich dynamic behaviour; 
SQUIDs and SQUID oligomers often reveal multistability, tunability, chaotic 
behavior, and complex bifurcation structures \cite{Hizanidis2018,Shena2020}. 
Furthermore, relatively large SQUID arrays in one and two dimensions support 
various types of spatially inhomogeneous states such as chimera states
\cite{Lazarides2015b,Hizanidis2016a}, spatially localized states 
of the discrete breather type \cite{Lazarides2017,Lazarides2018a}, 
as well as patterned (Turing-like) states \cite{Hizanidis2020}.  
A direct imaging technique of low-temperature Laser Scanning Microscopy (LSM) 
was applied to visualize spatially localized excitations in coupled rf SQUID 
arrays forming a magnetic meta-surface for electromagnetic wave propagation 
\cite{Zhuravel2019}. Often, rf SQUIDs are packed closely together side-by-side 
with substantial long-range (dipole-dipole) mutual inductance of the SQUID loops 
due to their close lateral proximity. The coupling between any two SQUIDs in an 
array in that case is therefore nonlocal, and falls-off {\em approximately} as 
the inverse cube of their center-to-center distance. 
It has been shown that nonlocal coupling significantly affects the resonance 
patterns in closely packed SQUID arrays \cite{Cai2024}. 

Such nonlinear and discrete systems are known to support localized modes of the 
discrete breather (DB) type. These modes are genuine nonlinear excitations that 
oscillate for long times both in conservative and driven-dissipative discrete 
and nonlinear systems 
\cite{Johansson1995,Flach1995,Rasmussen1997,Flach1998b,Eleftheriou2000,
Kevrekidis2001b,Schuster2001,Gorbach2005,Eleftheriou2008,Lazarides2008b,
Kominis2010,Lazarides2013a,Johansson2015,Bel2018,Dai2021, Naumov2023}; their 
very existence has been proposed already in 1988 \cite{Sievers1988}. 
Since then, a large volume of analytical and numerical investigations have 
explored the emergence of DBs in a variety of discrete nonlinear systems. 
Rigorous proofs of their existence have been provided for both Hamiltonian and 
dissipative systems \cite{Mackay1994,Aubry1997}, and numerical algorithms for 
their construction have been designed \cite{Marin2001,Zueco2005}. 
These modes appear spontaneously in a lattice as result of fluctuations 
\cite{Tsironis1996,Rasmussen2000} or disorder \cite{Rasmussen1999}, or by purely 
deterministic mechanisms \cite{Hennig2005}. Experimental observations of DBs have 
been made in solid state mixed-valence transition metal complexes \cite{Swanson1999},
quasi one-dimensional antiferromagnetic chains \cite{Schwarz1999}, arrays of 
Josephson junctions \cite{Schuster2004}, optical waveguide systems 
\cite{Eisenberg1998}, proteins \cite{Edler2004}, and micromechanical cantilever 
arrays \cite{Sato2007}. DBs modify system properties such as lattice thermodynamics, 
and often introduce nondispersive energy transport \cite{Kopidakis2001}. 
In the past few decades, DB investigation has attracted great interest in the 
scientific community and their connections to various  physical and engineering 
applications has been discussed 
\cite{Bountis2003,Bergamin2003,Bergamin2003,Campbell2004,Flach2008a}.

Here, we investigate localization due to nonlinearity in a Hamiltonian SQUID 
array/metamaterial, in which the total energy is conserved. The system is free
from dissipation and periodic forcing; however, a dc (constant) flux bias is 
allowed to be applied uniformly along the metamaterial/array. The emergence of
nonlinear localized modes of the DB type is then investigated through 
initialization with a single-site excitation in the middle of the array.
The existence of DBs in that system has been largely overlooked in literature,
and even the existence of dissipative DBs, i.e., the existence of DBs in 
periodically driven SQUID metamaterials in the presence of dissipation, has 
been only rarely addressed \cite{Lazarides2008a,Lazarides2012}.
Our simulations reveal a new type of localized mode of the DB type which is 
{\em asynchronous} in the sense that the SQUID at the central DB site oscillates
at a frequency different than that of the other SQUIDs in the metamaterial/array.
The SQUID at the central DB site oscillates at a frequency below the linear 
frequency band and it depends on the amplitude of the oscillation. To the 
contrary, the rest of the SQUIDs (i.e., the "background" SQUIDs) oscillate at 
frequencies within the linear frequency band. This situation is reminiscent of 
that of the existence of localized breathing modes in lattices with defects 
\cite{Theocharis2009} and is discussed in later Sections.

In the next Section (Section II), the model equations of the SQUID metamaterial 
are presented, and the linear and weakly nonlinear frequency dispersion relation 
is obtained. 
In Section III, we explore the existence of localized DB modes as a function of 
the model parameters using a standard localization measure.
In Section IV we present Fourier power spectrums obtained from time-series of 
the evolution of certain SQUIDs and discuss their oscillation frequencies.
The conclusions and further discussion are given in the last Section (Section V).

\section{Dynamic Equations and Frequency Dispersion}
Consider $N$ identical SQUIDs arranged on a one-dimensional array subject to a 
uniformly applied dc flux bias $\Phi_{dc}$ threading perpendicularly the SQUIDs' 
loops. The SQUIDs are coupled together weakly through dipole-dipole magnetic 
forces that fall-off roughly as the inverse cube of their center-to-center 
distance. Hence, the magnetic flux $\Phi_n$ threading the loop of the $n-$th 
SQUID in the array is
\begin{equation}
\label{eq01}
  \Phi_n =\Phi_{dc} +L\, I_n +L\, \sum_{m\neq n} \lambda_{|m-n|} I_m ,
\end{equation}
\begin{figure}[!h]
\includegraphics[angle=0, width=0.95 \linewidth]{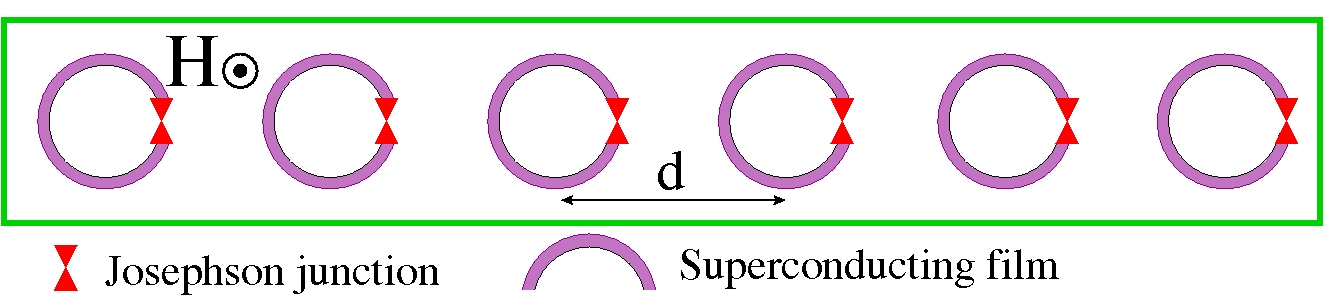} 
\caption{(Color online)
Schematic of a one-dimensional SQUID metamaterial in a planer configuration.
}
\label{fig02}
\end{figure}
where the indices $n$ and $m$ run from $1$ to $N$, $\lambda_{|m-n|} =M_{|m-n|}/L$ 
is the dimensionless coupling coefficient between the SQUIDs at the sites $m$ 
and $n$, with $M_{|m-n|}$ being their corresponding mutual inductance, and
\begin{eqnarray}
\label{eq02}
    -I_n = C\frac{d^2\Phi_n}{dt^2} +\frac{1}{R} \frac{d\Phi_n}{dt} 
          +I_c\, \sin\left( 2\pi\frac{\Phi_n}{\Phi_0} \right) 
\end{eqnarray}
is the current in each SQUID given by the resistively and capacitively shunted 
junction (RCSJ) model \cite{Likharev1986}, with $\Phi_0$ and $I_c$ being the flux 
quantum and the critical current of the Josephson junctions, respectively.  
Within the RCSJ model framework, $R$, $C$, and $L$ are the resistance, 
capacitance, and self-inductance of the SQUIDs' equivalent circuit,respectively 
(shown in Fig. \ref{fig01}b). Combination of Eqs. (\ref{eq01}) and (\ref{eq02}) 
gives
\begin{eqnarray}
\label{eq03}
  C\frac{d^2\Phi_n}{dt^2} +\frac{1}{R} \frac{d\Phi_n}{dt}
    +\frac{1}{L} \sum_{m=1}^N  \left( {\bf \hat{\Lambda}}^{-1} \right)_{nm} 
         \left[ \Phi_m -\Phi_{dc} \right] 
   \nonumber \\
   +I_c\, \sin\left(2\pi\frac{\Phi_n}{\Phi_0} \right) =0 ,
\end{eqnarray}
where ${\bf \hat{\Lambda}}^{-1}$ is the inverse of the $N\times N$ coupling 
matrix 
\begin{eqnarray}
\label{eq04}
  {\bf \hat{\Lambda}} = \left\{ \begin{array}{ll}
     1, & \mbox{if $m= n$}; \\
    \lambda_{|m-n|} =\lambda \, |m-n|^{-3}, & \mbox{if $m\neq n$},\end{array} 
    \right.    
\end{eqnarray}
with $\lambda$ being the coupling coefficient between nearest neighboring SQUIDs.

For sufficient spatial separation between SQUIDs, the system dynamics are 
dominated by nearest-neighbor coupling. In that case, only the terms with 
$m=n$ and $m=n \pm 1$ remain in the sum in Eq. (\ref{eq03}) which becomes
\begin{eqnarray}
\label{eq05}
  C \frac{d^2\Phi_n}{dt^2} +\frac{1}{R} \frac{d\Phi_n}{dt}
     -\frac{1}{L} \lambda 
    \left[ \Phi_{n+1} +\Phi_{n-1} -2\Phi_{dc} \right]
   \nonumber \\
   -\frac{1}{L} \Phi_{dc} +I_c\, \sin\left(2\pi\frac{\Phi_n}{\Phi_0}\right) =0,
\end{eqnarray}
where we used that $\left( {\bf \hat{\Lambda}}^{-1} \right)_{n,n} \simeq 1$ and
$\left( {\bf \hat{\Lambda}}^{-1} \right)_{n,n\pm 1} \simeq -\lambda$ 
(for the details of the derivation see Ref. \cite{Lazarides2018b}). 
Note that higher order terms in $\lambda$ are neglected in consistency with the 
nearest-neighbor coupling approximation ($\propto \lambda$).

Using the normalizing relations
\begin{equation}
\label{eq06}
    \tau=\omega_{LC} t, \qquad \phi =\frac{\Phi}{\Phi_0}, 
\end{equation}
with $\omega_{LC} =1/\sqrt{L C}$, and letting $R \rightarrow +\infty$, the 
normalized form of Eq. (\ref{eq05}) is obtained as
\begin{equation}
\label{eq07}
  \ddot{\phi}_n +\phi_n +\beta \sin\left( 2\pi \phi_n \right)
    =\lambda \left( \phi_{n+1} +\phi_{n-1} \right) +(1-2\lambda) \phi_{dc},
\end{equation}
where 
\begin{equation}
\label{eq08}
   \beta=\frac{I_c L}{\Phi_0} =\frac{\beta_L}{2\pi}
\end{equation}
is the rescaled SQUID parameter, and the overdots denote differentiation with 
respect to the normalized temporal variable $\tau$. 
\begin{figure}[!h]
\includegraphics[angle=0, width=0.95 \linewidth]{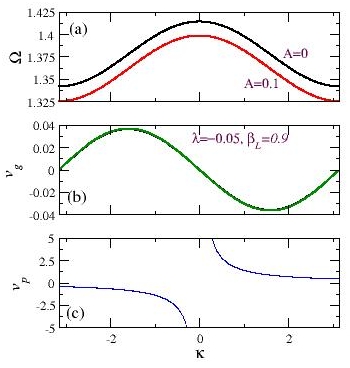} 
\caption{(Color online)
   (a) Linear (black curve) and  nonlinear (red curve) frequency dispersion of 
    the SQUID metamaterial as a function of the normalized wavenumber $\kappa$,
    plotted using Eqs. (\ref{eq11}) and (\ref{eq15}), respectively, for 
    $\lambda=-0.05$, $\beta_L =0.9$, and $A=0.1$. 
   (b) The corresponding group velocity dispersion $v_g$ as a function of $\kappa$, 
   and 
   (c) the corresponding phase velocity dispersion $v_p$ as a function of $\kappa$.
\label{fig03}
}
\end{figure}

Using the approximation $\beta \sin(2\pi \phi_n) \simeq \beta_L \phi_n$ and 
setting $\phi_{dc} =0$ in Eq. (\ref{eq08}, we obtain 
\begin{eqnarray}
\label{eq09}
  \ddot{\phi}_n +\Omega_{SQ}^2 \phi_n =\lambda \left( \phi_{n+1} +\phi_{n-1} \right)  
\end{eqnarray}
where $\Omega_{SQ} =\sqrt{ 1 +\beta_L }$. By substituting the plane wave 
solution
\begin{equation}
\label{eq10}
   \phi_n =A\, e^{i (\kappa n -\Omega \tau)},
\end{equation}
where $\kappa$ and $\Omega$ is the (dimensionless) wavenumber and frequency,
respectively, in Eq. (\ref{eq09}) we obtain the linear frequency dispersion 
relation
\begin{equation}
\label{eq11}
   \Omega(\kappa) =\sqrt{ \Omega_{SQ}^2 -2 \lambda \cos(\kappa) }.
\end{equation}
The above relation implies a narrow band with lower and upper edges equal to
\begin{equation}
\label{eq11.2}
   \Omega_{min} =\sqrt{\Omega_{SQ}^2 +2 \lambda} ~~\mbox{and}~~ 
   \Omega_{max} =\sqrt{\Omega_{SQ}^2 -2 \lambda},
\end{equation}
respectively ($\lambda < 0$). Since $\Omega_{SQ}$ is of order $1$ and 
$\lambda \ll 1$, the bandwidth $\Delta \Omega =\Omega_{max} -\Omega_{min}$
can be approximated by
\begin{equation}
\label{eq11.3}
   \Delta \Omega \simeq \frac{2 |\lambda|}{\Omega_{SQ}} =
        \frac{2 |\lambda|}{\sqrt{1 +\beta_L}}.
\end{equation}
\begin{figure}[!t]
\includegraphics[angle=0, width=0.95 \linewidth]{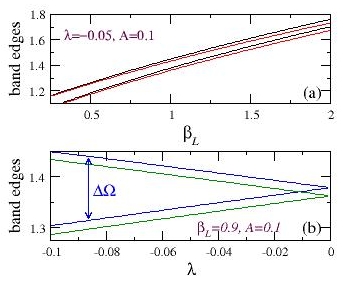} 
\caption{(Color online)
   (a) Linear (black curves) and nonlinear (red curves, $A =0.1$) frequency band 
       edges as a function of the SQUID parameter $\beta_L$ for $\lambda =-0.05$. 
   (b) Linear (blue curves) and nonlinear (green curves, $A =0.1$) frequency band 
       edges as a function of the coupling coefficient $\lambda$ for $\beta_L =0.9$.
}
\label{fig04}
\end{figure}

In the weak nonlinearity regime, the term $\beta \sin\left( 2\pi \phi_n \right)$ 
in Eq. (\ref{eq07}) can be approximated as 
\begin{equation}
\label{eq12}
   \beta \sin(2\pi \phi_n) \simeq \beta_L \phi_n -\chi \phi_n^3,
\end{equation}
where $\chi =2\pi^2 \beta_L /3$ is the nonlinearity coefficient. Hence, the 
normalized equation for the fluxes through the SQUID loops Eq.(\ref{eq07}) 
becomes ($\phi_{dc} =0$)
\begin{equation}
\label{eq13}
  \ddot{\phi}_n +\Omega_{SQ}^2 \phi_n -\chi \phi_n^3
    =\lambda \left( \phi_{n+1} +\phi_{n-1} \right) 
\end{equation}
Then, by substituting 
\begin{equation}
\label{eq14}
  \phi_n(\tau) =A e^{i \theta_n} +A^* e^{-i \theta_n}
\end{equation}
where $A^*$ is the complex conjugate of $A$ and $\theta_n =\kappa n -\Omega \tau$,
into Eq. (\ref{eq13}) and by using the rotating wave approximation (RWA) we obtain
the nonlinear frequency dispersion relation
\begin{equation}
\label{eq15}
  \Omega(\kappa) =\sqrt{ \Omega_{SQ}^2 -2 \lambda \cos(\kappa) -3 \chi |A|^2 }, 
\end{equation}
with $|A|$ being the half-amplitude of the linear wave in Eq. (\ref{eq14}).
Equation (\ref{eq15}) holds for $|A| < 0.32$, although in practice we should 
limit the values of $|A|$ up to $0.2$. This is however consistent with the 
weak nonlinearity approximation.

Typical linear and nonlinear frequency dispersions are shown in Fig. \ref{fig03}(a),
for SQUID parameter $\beta_L =0.9$ and coupling coefficient $\lambda =-0.05$.
The frequency bands are rather narrow due to the weak coupling between SQUIDs.
As it apparent from the figure, the effect of weak nonlinearity is to shift the
whole band downward by an amount that depends on $A$ and the nonlinearity 
coefficient $\chi$. The corresponding group and phase velocities $v_g$ and $v_p$,
respectively, are shown in Figs. \ref{fig03}(b) and (c) for the linear case.
Note that they have opposite sign everywhere along $\kappa$, a feature that is
characteristic to a (backward-wave) metamaterial.
\begin{figure}[!h]
\includegraphics[angle=0, width=0.95 \linewidth]{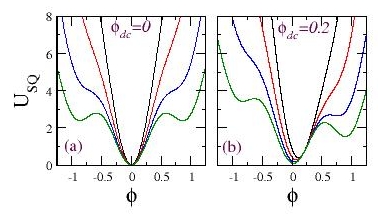} 
\caption{(Color online)
   The SQUID (on-site) potential $U_{SQ}$ as a function of the flux $\phi$ 
   through the SQUID loop for (a) $\phi_{dc} =0$; (b) $\phi_{dc} =0.2$.
   In both figures, the SQUID parameter takes the values $\beta_L =0.9$ (black 
   curves), $\beta_L =1.8$ (red curves), $\beta_L =3.6$ (blue curves), and
   $\beta_L =7.2$ (green curves).
}
\label{fig05}
\end{figure}

The edges of the linear and nonlinear frequency bands are shown in Figs. 
\ref{fig04}(a) and (b) as a function of $\beta_L$ and $\lambda$, respectively. 
In Fig. \ref{fig04}(a), the band edges in the linear regime increase almost 
linearly as a function of $\beta_L$ for fixed $\lambda =-0.05$ (black curves), 
while the bandwidth remains almost constant. In Fig. \ref{fig04}(b), the upper 
(lower) band edge in the linear regime increases (decreases) almost linearly as 
a function of the magnitude of $\lambda$, while the mid-band frequency remains 
constant and the bandwidth increases for fixed $\beta_L =0.9$ (blue curves). 
The corresponding band edges in the weakly nonlinear regime are shown as red and 
green curves in Figs. \ref{fig04}(a) and (b), respectively, are merely downshifted 
with respect to the band edges in the linear regime.

As can be readily observed from Eq. (\ref{eq07}), the (sinusoidal) nonlinearity
is local (on-site). Thus, an on-site potential can be assigned to any of the 
SQUIDs in the array. Let $\lambda$ be zero in Eq. (\ref{eq07}) to obtain 
\begin{equation}
\label{eq16}
  \ddot{\phi}_n +\phi_n +\beta \sin\left( 2\pi \phi_n \right) =\phi_{dc},
\end{equation} 
which is a Newtonian equation of the form (the subscript $n$ is dropped)
$\ddot{\phi} =-\partial U_{SQ} /\partial \phi$, where $U_{SQ}$ is the (single)
SQUID potential. Then, from Eq. (\ref{eq16}) we get that 
\begin{equation}
\label{eq17}
  U_{SQ}(\phi) =\frac{1}{2} \left( \phi -\phi_{dc} \right)^2 
                -\frac{\beta}{2\pi} \cos\left( 2\pi \phi \right).
\end{equation}
The SQUID potential $U_{SQ}$ is shown in Fig. \ref{fig05} for four different 
values of $\beta_L$ and two values of $\phi_{dc}$. In Fig. \ref{fig05}(a), for 
$\phi_{dc} =0$, we observe how the SQUID potential varies as a function of 
$\beta_L$; for $\beta_L =0.9 < 1$ the potential is almost parabolic. However, the 
potential becomes a corrugated parabolic one for $\beta_L >1$ and even acquires 
more minimums for large enough $\beta_L$ (in the present case, for $\beta_L =7.2$, 
green curve). 
In Fig. \ref{fig05}(b), for $\phi_{dc} =0.2$, we observe how the SQUID potential 
develops an asymmetry due to the dc bias flux. Besides that, the dc bias flux 
shifts the minimums of the potential away from zero. This effect is more 
pronounced for small values of $\beta_L$. 
\begin{figure}[!h]
\includegraphics[angle=0, width=0.95 \linewidth]{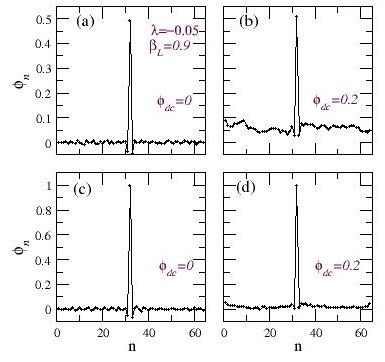} 
\caption{(Color online)
   Discrete breather profiles for $N=64$, $\lambda =-0.05$, $\beta_L =0.9$, and
   (a) $\phi_{in} =0.5$ and $\phi_{dc} =0$; (b) $\phi_{in} =0.5$ and $\phi_{dc} =0.2$;
   (c) $\phi_{in} =1.0$ and $\phi_{dc} =0$; (b) $\phi_{in} =1.0$ and $\phi_{dc} =0.2$.
}
\label{fig06}
\end{figure}

\section{Nonlinear Localization -- Discrete Breathers}
Equations (\ref{eq07}) are integrated in time using a sixth-order Runge-Kutta 
algorithm with fixed time-step $\Delta \tau$ \cite{Sarafyan1972}, typically 
$0.05$ and $0.02$ (for the Fourier power spectrums presented in Section IV), 
for $N =64$ (number of SQUIDs in the array), and free-end boundary conditions to 
account for a finite metamaterial. Decreasing the time-step size yielded 
identical results, confirming temporal convergence and numerical stability. 
The system is initialized at $\tau=0$ with a single-site excitation of the form
\begin{eqnarray}
\label{eq18}
  \phi_n = \left\{ \begin{array}{ll}
     \phi_{in}, & \mbox{if $n= n_b =N/2$}; \\
      0         & \mbox{if $n\neq n_b$},\end{array} \right. ~~
  \dot{\phi}_n =0, \mbox{for any $n$}, 
\end{eqnarray}
where $\phi_{in}$ is the amplitude of the single-site initial excitation and 
$n=1,2, ... ,N$. 
\begin{figure}[!h]
\includegraphics[angle=0, width=0.95 \linewidth]{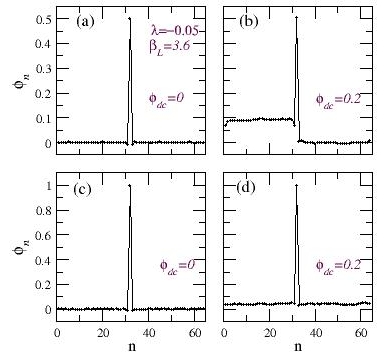} 
\caption{(Color online)
   Discrete breather profiles for $N=64$, $\lambda =-0.05$, $\beta_L =3.6$, and
   (a) $\phi_{in} =0.5$ and $\phi_{dc} =0$; (b) $\phi_{in} =0.5$ and $\phi_{dc} =0.2$;
   (c) $\phi_{in} =1.0$ and $\phi_{dc} =0$; (b) $\phi_{in} =1.0$ and $\phi_{dc} =0.2$.
}
\label{fig07}
\end{figure}

Typical profiles of highly localized discrete breathers are shown in Figs. 
\ref{fig06} and \ref{fig07} at a time-instant at which $\phi_{n_b}(\tau)$ is 
maximum (i.e., equal to the actual breather amplitude, $\phi_{br}$). They are 
obtained after integrating Eq. (\ref{eq07}) in time for $50,000$ time units which 
correspond to approximately $8,000-16,000$ breather periods (depending on the 
actual value of the breather period). Longer integration times up to $1,000,000$ 
time units were also used, which yielded practically identical results. In Figs. 
\ref{fig06}(a) and (b), the system is initialized with a single-site excitation of 
amplitude $\phi_{in} =0.5$ for flux bias $\phi_{dc} =0$ and $0.2$, respectively. 
The obtained profiles feature highly localized breathers oscillating on a low 
amplitude oscillating background. This is because the single-site initial excitation 
is not an exact solution of Eq. (\ref{eq07}). Thus, when the breather is formed, the 
residual energy is transported through the array causing the SQUIDs in the background 
to oscillate with low amplitude at frequencies within the linear frequency band. In 
Figs. \ref{fig06}(c) and (d), the system is initialized with $\phi_{in} =1$ for flux 
bias $\phi_{dc} =0$ and $0.2$, respectively. The obtained profiles are very similar 
to these shown in \ref{fig06}(a) and (b). Note that when the metamaterial is subject 
to a flux bias $\phi _{dc}$ (as in Figs. \ref{fig06}(b) and (d)), the background does 
not average to zero, unlike the case without a flux bias. This can be seen more 
clearly in Fig. \ref{fig06}(b) in which the background averages at around $0.05$.

The breather profiles in Fig. \ref{fig07} are obtained using identical parameters 
to those in Fig. \ref{fig06}, with the exception that $\beta_L =3.6$. 
The observations made for the breather profiles in Fig. \ref{fig06} also apply 
to the breather profiles in Fig. \ref{fig07}. A new important feature here is
that the background in Fig. \ref{fig07}(b) averages out to two distinct levels.
\begin{figure}[!h]
\includegraphics[angle=0, width=0.95 \linewidth]{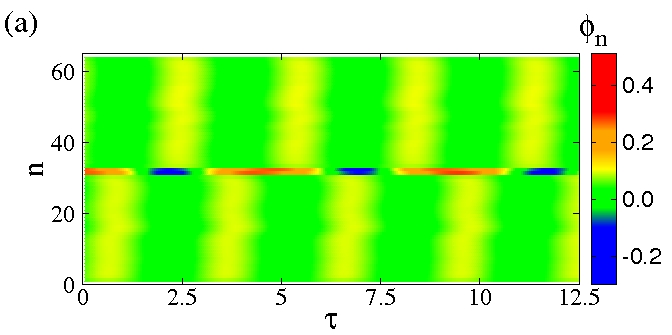} \\
\includegraphics[angle=0, width=0.95 \linewidth]{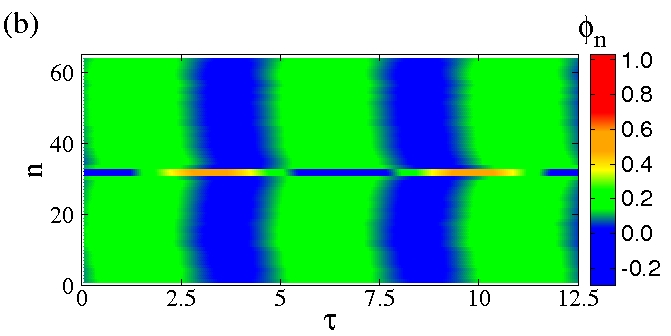}
\caption{(Color online)
   The spatiotemporal evolution of two discrete breathers is shown as a map of 
   $\phi_n$ on the site number - normalized time ($n - \tau$) plane for $N=64$,
   $\lambda=-0.05$, $\phi_{dc} =0.2$, and
   (a) $\phi_{in} =0.5$ and $\beta_L =3.6$;
   (b) $\phi_{in} =1.0$ and $\beta_L =0.9$.
\label{fig08}
}
\end{figure}

To better comprehend this effect, we map the spatiotemporal evolution of two DBs
on the $n - \tau$ plane in Fig. \ref{fig08}. For Fig. \ref{fig08}(a) and 
\ref{fig08}(b) the same parameters as in Fig. \ref{fig07}(b) and \ref{fig06}(d)
are used. As it can be observed in Fig. \ref{fig08}(a), the oscillations of two
SQUIDs left and right from the central DB site at $n=n_b =N/2$) have opposite 
phases. One can also observe that the oscillation of the SQUID at at the central 
DB site are not symmetrical with respect to the background. It also seems that 
there is a phase difference in the oscillations between consecutive SQUIDs in the 
background which increases linearly from the SQUID next to the central DB site 
toward the ends of the array. To the contrary, Fig. \ref{fig08}(b) reveals that 
the two SQUID oscillations left and right from the SQUID at the central DB site 
are in phase. The spatiotemporal evolution of these DBs is mapped for 
approximately two periods of oscillation. During this time interval, the 
difference in the frequency of oscillations between the SQUID at the central 
breather site and the SQUIDs in the background is not easily discernible. This 
issue is clarified in the next Section in which we examine the Fourier power 
spectrums of the oscillations.
\begin{figure*}[!t]
\includegraphics[angle=0, width=0.4 \linewidth]{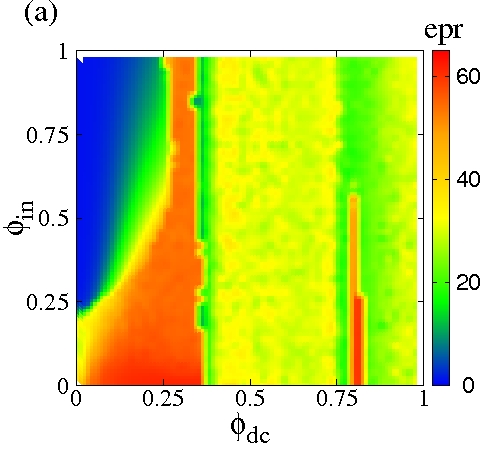}
\includegraphics[angle=0, width=0.4 \linewidth]{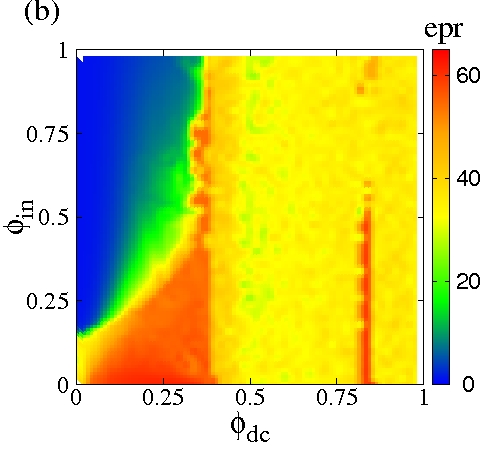} \\
\includegraphics[angle=0, width=0.4 \linewidth]{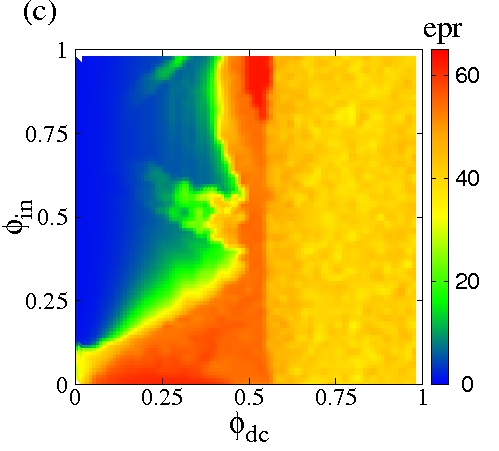}
\includegraphics[angle=0, width=0.4 \linewidth]{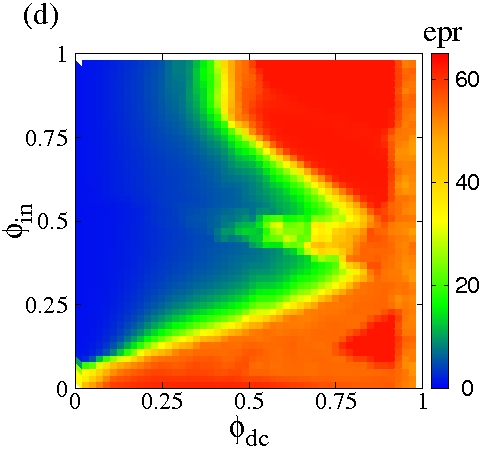}
\caption{(Color online)
   Two dimensional maps of the energetic participation ratio ($epr$) as a 
   function of the initial single-site excitation amplitude $\phi_{in}$ and the 
   flux bias $\phi_{dc}$, for $N=64$, $\lambda =-0.05$, and
   (a) $\beta_L =0.9$; (b) $\beta_L =1.8$; (c) $\beta_L =3.6$; (d) $\beta_L =7.2$.
   Blue color indicates areas in which highly localized breathers exist.
}
\label{fig09}
\end{figure*}

For a parametric investigation of nonlinear localization in a SQUID metamaterial
/array, we need an appropriate measure. In that case, in which a conserved 
Hamiltonian function is available for the system, we use the energetic 
participation ratio ($epr$), whose calculation is described below. Equation 
(\ref{eq07}) can be obtained through Hamilton's equations with the (symmetrized) 
Hamiltonian function
\begin{eqnarray}
\label{eq19}
   H =\sum_n H_n =\sum_n \left\{ 
      \frac{1}{2} \dot{\phi}_n^2 +\frac{1}{2} \left( \phi_n -\phi_{dc} \right)^2
      -\frac{\beta}{2\pi} \sin(2\pi \phi_n) 
      \right.
      \nonumber \\
      \left.
      -\frac{\lambda}{2} \left[
      (\phi_n -\phi_{dc}) (\phi_{n-1} -\phi_{dc}) +(\phi_{n+1} -\phi_{dc}) (\phi_n -\phi_{dc})
      \right] \right\}
      \nonumber \\
\end{eqnarray}
where $H_n$ is the Hamiltonian density, whose definition is obvious from the 
above equation. The Hamiltonian $H$ and the associated Hamiltonian density $H_n$ 
correspond to the total energy and the energy density of the SQUID metamaterial 
in the presence of flux bias $\phi_{dc}$. The energetic participation ratio $epr$ 
is defined as
\begin{equation}
\label{eq20}
  epr =\left[ \frac{\sum_n \epsilon_n^2}{\left( \sum_n \epsilon_n \right)^2 }
  \right]^{-1},
\end{equation} 
where 
\begin{equation}
\label{eq21}
  \epsilon_n =\frac{H_{n,M}}{H_M},
\end{equation} 
with $H_{n,M}$ and $H_M$ being the modified Hamiltonian density and (total) 
Hamiltonian, respectively. The latter are obtained by subtracting the minimum
of $H_n$, $min\{H_n\}$, from $H_n$, i.e., $H_{n,M} =H_n -min\{H_n\}$ shifting 
the energies of all SQUID oscillators upward by the same amount. Then, we add 
the modified energies of each SQUID oscillator in the array to obtain the 
modified Hamiltonian $H_M =\sum_n H_{n,M}$. The energetic participation ratio 
thus defined takes values from $1$ to $N$, which correspond to a highly localized 
mode (DB) where almost all the energy is localized to the SQUID at the 
central DB site, and a completely delocalized (extended) mode where the energy 
is equipartitioned across the SQUID oscillators.
\begin{figure}[!h]
\includegraphics[angle=0, width=0.95 \linewidth]{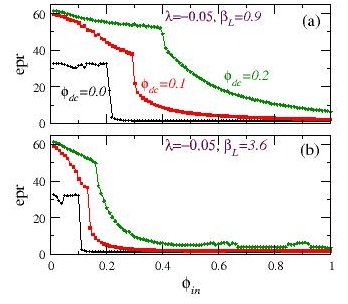} 
\caption{(Color online)
   The energetic participation ratio ($epr$) as a function of the initial 
   single-site excitation amplitude $\phi_{in}$, for $N=64$, $\lambda =-0.05$,
   and (a) $\beta_L =0.9$; (b) $\beta_L =3.6$. In both (a) and (b) the flux bias 
   is $\phi_{dc} =0$ (black curves), $\phi_{dc} =0.1$ (red curves), and
   $\phi_{dc} =0.2$ (green curves),
}
\label{fig10}
\end{figure}

In Fig. \ref{fig09}, the energetic participation ratio $epr$ is mapped on the 
single-site excitation amplitude ($\phi_{in}$) - flux bias ($\phi_{dc}$) plane,
for four values of the SQUID parameter $\beta_L$ (see caption). In these figures,
localized modes of the DB type exist in the blue areas. As it can be observed, 
stable DBs exist when the single-site initial excitation amplitude $\phi_{in}$ 
exceeds a threshold value. Generally speaking, increasing flux bias $\phi_{dc}$ 
increases that threshold value in all the four cases shown.
We also observe that nonlinear localization is favored by larger values of 
$\beta_L$. (Note the increase in size of the blue areas from Fig. \ref{fig09}(a) 
to \ref{fig09}(d). To calculate the $epr$ at each point on the maps in Fig. 
\ref{fig09}, Eq. (\ref{eq07}) is first integrated for $50,000$ time units to 
eliminate transient dynamics and ensure the stability of a localized mode. 
Subsequently, the system is evolved over two to three SQUID oscillation periods, 
recording instantaneous $epr$ values at each time-step. The final reported $epr$ 
corresponds to the temporal average over this time interval.

The transition from a delocalized (extended) to a localized mode is not smooth, 
but it is accompanied with a jump from high to low values of $epr$. This is not 
however easily discernible on the maps; for that reason we present in Fig. 
\ref{fig10} the $epr$ dependence on the single-site excitation amplitude $\phi_{in}$
for a few selected values of $\phi_{dc}$ and $\beta_L$. In Fig. \ref{fig10}(a) 
(for $\beta_L =0.9$), the three curves are obtained under different flux bias. 
For $\phi_{dc} =0$, there is an abrupt transition at about $\phi_{in} =0.21$ 
(black curve); in this case, the $epr$ decreases from approximately $32$ to almost 
$1$ when $\phi_{in}$ crosses the threshold. For $\phi_{dc} =0.1$ and $0.2$ (red 
and green curves, respectively) the transition is smoother but still visible, 
signified by a jump at the transition point (at $\phi_{in} =0.3$ and 
$\phi_{in} =0.4$, respectively). Qualitatively similar results are obtained for 
larger $\beta_L$ ($3.6$), as it is shown in Fig. \ref{fig10}(b). For larger 
$\beta_L$, lower threshold values are required for the transition from a 
delocalized (extended) to a localized mode to occur. The corresponding threshold 
values for $\phi_{in}$ in this case are, respectively, $\phi_{in} =0.11$, $0.14$, 
and $0.17$ for $\phi_{dc} =0$, $0.1$, and $0.2$.
\begin{figure}[!h]
\includegraphics[angle=0, width=0.95 \linewidth]{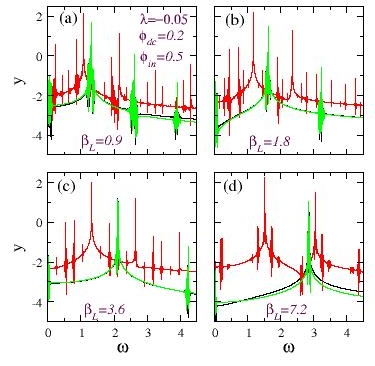} 
\caption{(Color online)
   The decimal logarithm of the Fourier power spectrums of $\phi_{N/4}(\tau)$, 
   $\phi_{n_b}(\tau)$, and $\phi_{N/4}(\tau)$, as a function of the Fourier 
   frequencies $\omega$, obtained from the time-series data for the oscillating 
   SQUIDs at $n =N/4$ (black curve), $n =n_b =N/2$ (red curve), $n =3N/4$ 
   (green curve), for $N=64$, $\lambda =-0.05$, $\phi_{dc} =0.2$, $\phi_{in} =0.5$, 
   and
   (a) $\beta_L =0.9$; (b) $\beta_L =1.8$; (c) $\beta_L =3.6$; (d) $\beta_L =7.2$.
   The symbol $y$ stands for $y =\log[PS(\phi_{n}(\tau))]$, with $n$ taking the 
   values specified above.
}
\label{fig11}
\end{figure}

\section{Asynchronous Oscillations - Power Spectrums}
In this Section we investigate the frequencies of oscillations of the SQUIDs of
the metamaterial, using temporal Fourier power spectrums of time-series data 
corresponding to a few particular SQUID oscillators. In Fig. \ref{fig11}, the 
Fourier power spectrums for three SQUIDs are shown; the SQUID at the central 
DB site at $n =n_b =N/2$ (red curve), as well as two SQUIDs in the background, 
left and right from the SQUID at the central DB site, at $n=N/4$ (black curve)
and $3N/4$ (green curve). As it can be observed, the black and green curves are
mostly overlapping. Such Fourier power spectrums are shown for four values of 
the SQUID parameter $\beta_L$ for a moderately high single-site initial excitation 
amplitude $\phi_{in} =0.5$, $\lambda=-0.05$, and $\phi_{dc} =0.2$. In all these
figures, a second harmonic appears, due to the non-zero flux bias $\phi_{dc}$ that 
breaks the symmetry of the SQUID (on-site) potential. The Fourier power spectrum 
for the SQUID at the central DB site consists of a number of harmonics and linear 
combinations of them. The corresponding Fourier power spectrum for the SQUIDs at 
$n =N/4$ and $n =3N/4$ consist of only a few harmonics. It can be clearly observed
that the dominant frequencies of the oscillating SQUIDs at the central breather 
site and the two SQUIDs at the two background sites are different. Specifically, 
the dominant oscillation frequency of the SQUID at the central DB site is lower 
(i.e., below the linear frequency band) than the dominant frequency of the two 
SQUIDs located at two sites in the background. As it can be observed, the 
difference between the dominant frequencies increases with increasing $\beta_L$. 
\begin{figure}[!h]
\includegraphics[angle=0, width=0.95 \linewidth]{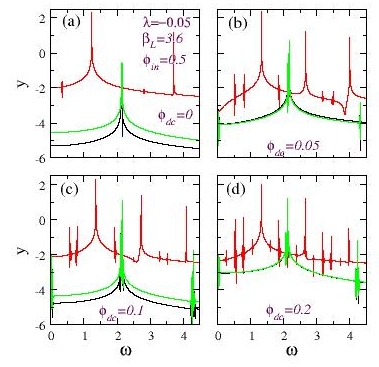} 
\caption{(Color online)
   The decimal logarithm of the Fourier power spectrums of $\phi_{N/4}(\tau)$, 
   $\phi_{n_b}(\tau)$, and $\phi_{N/4}(\tau)$, as a function of the Fourier 
   frequencies $\omega$, obtained from the time-series data for the oscillating 
   SQUIDs at $n =N/4$ (black curve), $n =n_b =N/2$ (red curve), $n =3N/4$ 
   (green curve), for $N=64$, $\lambda =-0.05$, $\beta_L =3.6$, $\phi_{in} =0.5$, 
   and
   (a) $\phi_{dc}=0$; (b) $\phi_{dc} =0.05$; (c) $\phi_{dc} =0.1$; (d) $\phi_{dc} =0.2$.
   The symbol $y$ stands for $y =\log[PS(\phi_{n}(\tau))]$, with $n$ taking the 
   values specified above.
}
\label{fig12}
\end{figure}

In Fig. \ref{fig12}, the temporal Fourier power spectrums of time-series data 
for the three above mentioned SQUIDs is shown for $\beta_L =3.6$ and four values 
of flux bias $\phi_{dc}$. The sharp peaks at the dominant frequency and its 
harmonics do not show any significant dependence of the flux bias $\phi_{dc}$. 
Still, the SQUID at the central breather site is oscillating with a frequency 
lower than that of the other two SQUIDs in the background, Actually, the 
frequencies of the SQUIDs in the background are limited to those within the 
linear frequency band. Note that there is no second harmonic in the spectrums 
for $\phi_{dc} =0$ (Fig. \ref{fig12}(a)).
\begin{figure}[!h]
\includegraphics[angle=0, width=0.95 \linewidth]{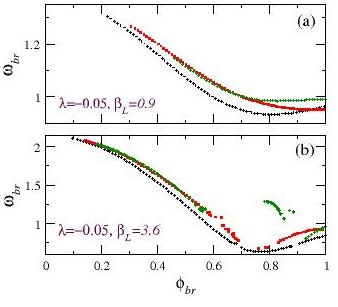} 
\caption{(Color online)
   The oscillation frequency of the SQUID at the central breather site at 
   $n =n_b =N/2$, $\omega_{br}$, as a function of the measured breather amplitude
   $\phi_{br}$ for $N=64$, $\lambda=-0.05$, and (a) $\beta_L =0.9$; 
   (b) $\beta_L =3.6$. In both (a) and (b) the black, red, and blue curves are
   obtained for $\phi_{dc}=0$, $0.1$, and $0.2$, respectively.
}
\label{fig13}
\end{figure}

Although the (dominant) frequencies of oscillations of the SQUIDS in the 
background are within the linear frequency band, the (dominant) oscillation 
frequency of the SQUID at the central DB site $\omega_{br}$ exhibits a strong 
dependence on the actual breather amplitude $\phi_{br}$. The latter is the DB 
amplitude measured from numerical data after Eq. (\ref{eq07}) is integrated for 
$50,000$ time units. Typical $\omega_{br} - \phi_{br}$ dependence for 
$\beta_L =0.9$ and $\beta_L =3.6$ are shown in Fig. \ref{fig13}(a) and (b), 
respectively, for three values of bias flux $\phi_{dc}$. In both figures, the 
black, red, and green points are obtained for $\phi_{dc} =0$, $0.1$, and $0.2$, 
respectively. As it should be expected, 
the highest DB frequencies are below the lower edge of the linear frequency band.
Using the linear frequency dispersion, Eq. (\ref{eq11}), the lower edge of the 
linear frequency band is at $\Omega_{min} =1.34$ and $\Omega_{min} =2.12$, 
respectively, for $\beta_L =0.9$ and $\beta_L =3.6$, respectively. As it can be 
observed from both figures, and for all curves shown, the $\omega_{br} - \phi_{br}$ 
relation is non-monotonic. Instead, the breather frequency $\omega_{br}$ decreases 
down to a minimum value with increasing $\phi_{br}$ (at approximately 
$\phi_{br} =0.8$ and $0.75$ for $\beta_L =0.9$ and $\beta_L =3.6$, respectively), 
and then increases with further increasing $\phi_{br}$. For fixed $\phi_{br}$, the 
DB frequencies for $\phi_{dc} =0$ are lower than those for $\phi_{dc} =0.1$ and 
$0.2$ (except in a small $\phi_{br}$ interval in Fig. \ref{fig13}(a) for 
$\phi_{br} > 0.9$).

As it is observed in Fig. \ref{fig13}(a), the $\omega_{br} - \phi_{br}$ curves 
for $\phi_{dc} =0.1$ and $0.2$ (red and green points, respectively) are the same 
for breather amplitude $\phi_{br} \simeq 0.75$. For larger $\phi_{br}$, the two 
curves diverge. In Fig. \ref{fig13}(b), the $\omega_{br} - \phi_{br}$ curves for 
$\phi_{dc} =0.1$ and $0.2$ (red and green points, respectively) are the same for 
breather amplitude $\phi_{br} \simeq 0.56$. For larger $\phi_{br}$, an isolated 
green branch appears around $\phi_{br} =0.8$, which is due to a breather family 
emerging at high breather amplitudes. Generally speaking, high values of 
$\beta_L$ increase complexity, and the analysis of such effects is beyond the 
scope of this paper.

\begin{figure}[!t]
\includegraphics[angle=0, width=0.8 \linewidth]{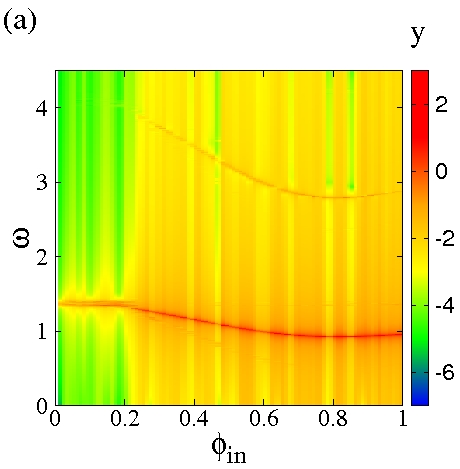} \\
\includegraphics[angle=0, width=0.8 \linewidth]{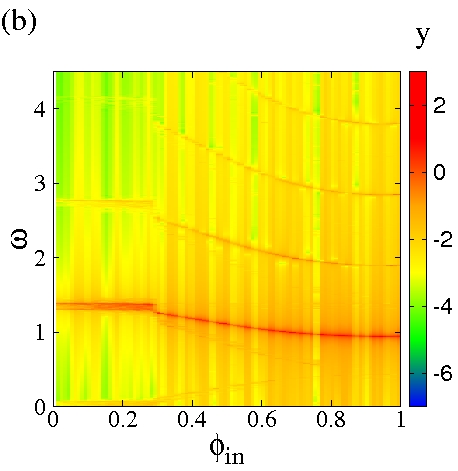} 
\caption{(Color online)
   The decimal logarithm of the power spectrums of $\phi_{n_b}(\tau)$, 
   $y =\log[PS(\phi_{n_b}(\tau))]$, mapped as a function of the single-site 
   initial excitation amplitude $\phi_{in}$ for $N=64$, $\lambda =-0.05$, 
   $\beta_L =0.9$, and (a) $\phi_{dc} =0$; (b) $\phi_{dc} =0.1$. 
}
\label{fig14}
\end{figure}

\begin{figure}[!h]
\includegraphics[angle=0, width=0.8 \linewidth]{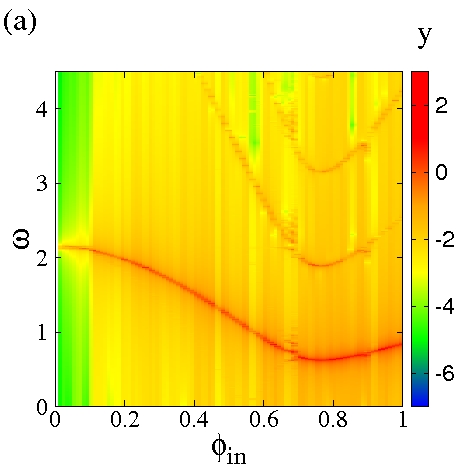} \\
\includegraphics[angle=0, width=0.8 \linewidth]{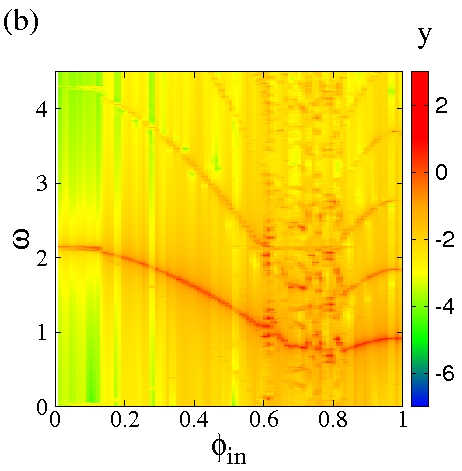} 
\caption{(Color online)
   The decimal logarithm of the Fourier power spectrums of $\phi_{n_b}(\tau)$, 
   $y =\log[PS(\phi_{n_b}(\tau))]$, mapped as a function of the single-site 
   initial excitation amplitude $\phi_{in}$ for $N=64$, $\lambda =-0.05$, 
   $\beta_L =3.6$, and (a) $\phi_{dc} =0$; (b) $\phi_{dc} =0.1$. 
}
\label{fig15}
\end{figure}

Better insights about the localization transition can be obtained from Figs. 
\ref{fig14} and \ref{fig15}, which show maps of the decimal logarithm of the 
power spectrums of $\phi_{n_b}(\tau)$, $y =\log[PS(\phi_{n_b}(\tau))]$, as a 
function of the single-site excitation amplitude $\phi_{in}$. Recall that 
$\phi_{n_b}$ is the flux oscillation amplitude of the SQUID at the central 
breather site, i.e., at $n =n_b =N/2$. The power spectrums are calculated from 
time-series data obtained after Eq. (\ref{eq07}) is integrated in time for 
$50,000$ time units to ensure the stability of the localized mode. The length 
of each time-series is $n_F =2^{20}$ time-steps, or 
$2^{20} \Delta t =2^{20} \times (0.02)=20,971.5$ time-units which implies a 
frequency resolution for the Fourier frequencies of 
$\Delta \omega \simeq 5\times 10^{-5}$. The time-series are Fourier transformed, 
and the power spectrums $PS(\phi_{n_b}(\tau))$ are obtained from 
\begin{equation}
\label{eq22}
   PS[\phi_{n_b}(\tau)]
    =\frac{1}{\sqrt{n_F}} \sqrt{ [Re(\bar{\phi}_{n_b})]^2 +[Im(\bar{\phi}_{n_b})]^2},
\end{equation} 
where $\bar{\phi}_{n_b}$ is the Fourier transform of $\phi_{n_b}(\tau)$.

In Fig. \ref{fig14}(a), for $\phi_{dc} =0$, we observe the variation of the 
dominant DB frequency as $\phi_{in}$ varies. For small $\phi_{in}$, no 
localization occurs and thus the dominant frequency follows a horizontal line 
segment (from $\phi_{in} =0$ to $0.21$). Note the considerable width of this 
segment, which is of the order of the linear frequency bandwidth. The latter can 
be calculated from Eq. (\ref{eq11}) for $\beta_L =0.9$ and $\lambda =-0.05$ to 
be $\Omega_{max} -\Omega_{min} \simeq 0.07$, with $\Omega_{min} =1.34$ and 
$\Omega_{max} =1.41$ being the lower and upper band edge, respectively.
For larger $\phi_{in}$ localization sets in, and the dominant frequency decreases 
with increasing $\phi_{in}$ until it reaches a minimum, and then starts 
increasing again with further increase of $\phi_{in}$. The dependence of the 
dominant breather frequency is more clearly seen for that case in Fig. 
\ref{fig13}(a) (black curve). 

Another important feature is the development of a third harmonic for values of
$\phi_{in}$ larger than the threshold $0.21$, which exhibits qualitatively the 
same dependence on $\phi_{in}$ as that of the dominant frequency (first harmonic).  
The same map is shown in Fig. \ref{fig14}(b), but for $\phi_{dc} =0.1$. Here, 
the threshold for localization is at $\phi_{in} =0.30$. For larger values of 
$\phi_{in}$ localization sets in, and the dominant frequency decreases with 
increasing $\phi_{in}$ until it reaches a minimum, and then increases again with 
further increasing $\phi_{in}$. Here, the development of not only the third but 
also even harmonics (second and fourth) is clearly visible. The appearance of 
the even harmonics in the spectrum is due to the symmetry-breaking action of the 
flux bias $\phi_{dc}$.

Figure \ref{fig15}, for $\beta_L =3.6$, presents qualitatively the same features 
as Fig. \ref{fig14}. However, the variation of the dominant frequency of the DB 
is much larger here. As it is observed from Fig. \ref{fig15}(a), the dominant DB 
frequency varies from $2.1$ (maximum value, for $\phi_{in}$ just above the
threshold) to $0.65$ (minimum value). Note that the localization thresholds here
are lower than those in Fig. \ref{fig14}, i.e., at $\phi_{in} =0.11$ and $0.14$ 
for $\phi_{dc} =0$ and $0.1$, respectively. The linear frequency bandwidth in 
this case is $\Omega_{max} -\Omega_{min} \simeq 0.05$, with $\Omega_{min} =2.12$ 
and $\Omega_{max} =2.17$ being the lower and upper band edge, respectively.

\section{Discussion and Concluding Remarks}
We investigated a finite, one-dimensional SQUID array/metamaterial subject to 
a spatially uniform dc flux bias with respect to its nonlinear localization 
properties. For that purpose, single-site initial excitations of varying 
amplitude were used along with free-end boundary conditions. The degree of 
localization is quantified by the energetic participation ratio $epr$ described 
in Section III. The system is modeled by the dimensionless Eq. (\ref{eq07}) which 
contains the coupling coefficient $\lambda$, the rescaled SQUID parameter $\beta$ 
and the externally applied flux bias $\phi_{dc}$, as parameters. The amplitude of 
the single-site initial excitation $\phi_{in}$ serves as a fourth parameter. Due 
to the nature of the coupling between SQUIDs, the magnitude of the (dimensionless) 
coupling coefficient $\lambda$ is much less than $1$ indicating strong 
discreteness and thus justifying the nearest-neighbor approximation adopted here. 
This parameter is kept fixed in this work, having the value $-0.05$. 

The value of second parameter, $\beta =\beta_L/(2\pi)$, is fixed by the 
as-fabricated loop geometry and junction characteristics of the SQUIDs. Here, we 
consider four values of $\beta_L$ and change continuously the flux bias 
$\phi_{dc}$ and/or the single-site excitation amplitude $\phi_{in}$. Larger 
$\beta_L$ favors nonlinear localization since $\beta =\beta_L/(2\pi)$ is the 
coefficient of the nonlinear term $\beta \sin(2\pi\phi_n$ in Eq. (\ref{eq07}); 
thus, larger $\beta_L$ implies stronger nonlinearity. On the other hand, larger 
$\phi_{dc}$ tends to prevent nonlinear localization. The single-site excitation 
amplitude $\phi_{in}$ is also an important parameter; larger $\phi_{in}$ triggers 
stronger nonlinear effects favoring localization. The transition from a 
delocalized/extended mode to a localized mode occurs when $\phi_{in}$ exceeds a 
certain threshold value that depends on the values of the other parameters. 
Similar results have been reported in the past for a standard discrete nonlinear 
Schr{\"o}dinger (DNLS) equation initially excited at a single site both 
numerically \cite{Eisenberg2002,Matuszewski2006} and experimentally 
\cite{Eisenberg2002}, in the context of optics. 

The calculated power spectrums presented in Section IV reveal that the SQUID at
the central site of the localized modes/breathers oscillates at a frequency 
which is different that those of the background sites. Specifically, the SQUID 
at the central breather site oscillates at a frequency $\omega_{br}$ below the 
lower edge of the linear band, while the background SQUIDs oscillate at a 
frequencies within the linear frequency. Such modes are reminiscent of localized 
breathing modes appearing in lattices of coupled oscillators with an impurity, 
see e.g., in Ref. \cite{Theocharis2009}. The dependence of the DB frequencies on 
the actual (i.e., the measured) breather amplitude $\phi_{br}$ is non-monotonic. 
Non-zero flux bias shifts the frequency-amplitude curves slightly upward. 

The appearance of breathers in the SQUID array is due to the well-known and 
investigated self-trapping phenomenon resulting from the interplay of 
discreteness and nonlinearity. When the single-site initial excitation amplitude 
exceeds a certain threshold, the excitation is capable to modify locally the 
physical properties of the lattice; this modification acts as a self-induced 
effective potential well \cite{Molina1993} which shifts the frequency of the 
SQUID oscillation at the initially excited site below the linear frequency band. 
Thus, the breather is decoupled from the linear modes and cannot transfer energy 
to them resulting in a dynamical self-trapping within the potential well it 
created.

To the best of our knowledge, this is the first time that an equation such as Eq. 
(\ref{eq07}) is investigated with respect to the existence of localized modes of
the discrete breather type. A natural extension of this work is to explore the
case of non-local coupling between SQUIDs, i.e., by using the normalized form of 
Eq. (\ref{eq03}) instead of Eq. (\ref{eq07}) for our simulations. Non-local 
coupling modifies the linear dispersion properties of DBs, it shifts their 
frequency spectrums, and changes the thresholds for their existence. In that 
context, it would be interesting to investigate how breather mobility and 
stability are modified if the SQUIDs are non-locally coupled.

\section*{Acknowledgments}
The author is grateful to Steven Anlage, Johanne Hizanidis, and Jingnan Cai for  
valuable discussions on SQUID metamaterials. 

\bibliography{BibTex-Library-05Jun2022}

\end{document}